\documentclass[a4paper,11pt]{article}
\pdfoutput=1 

\usepackage{jinstpub} 

\title{\boldmath Atom Interferometry for Dark Contents of the Vacuum Searches}

\author[a,1]{O. Burrow,\note{Present address: Department of Physics, John Anderson Building, University of Strathclyde, Glasgow, G4 0NG}}
\author[a]{A. Carroll,}
\author[b,c,2]{S. Chattopadhyay,\note{Previously at the Cockcroft Institute, UK, and Department of Physics, University of Liverpool, UK.}}
\author[a,3]{ J. Coleman,\note{Corresponding author.}}
\author[a, d]{G. Elertas,}
\author[a]{J. Heffer,}
\author[a]{C. Metelko,}
\author[a, d]{R. Moore,}
\author[a]{D. Morris,}
\author[e,4]{M. Perl,\note{Deceased.}}
\author[a]{J. Ralph,}
\author[a, d]{J. Tinsley}

\affiliation[a]{Department of Physics, University of Liverpool, Oliver Lodge Building, Oxford Street, Liverpool, L69 7ZE, UK}
\affiliation[b]{Fermi National Accelerator Laboratory, Wilson and Kirk Roads, Batavia, Illinois 60510, USA}
\affiliation[c]{Department of Physics, Northern Illinois University, IL 60115, USA}
\affiliation[d]{National Physical Laboratory, Teddington, Middlesex, TW11 0LW, UK}
\affiliation[e]{SLAC, National Accelerator Laboratory, Stanford University, Menlo Park, CA 94025,  USA}

\emailAdd{coleman@liverpool.ac.uk}

\abstract{A cold atom interferometer is being developed using $^{85}$Rb atoms towards a search for the dark contents of the vacuum, and as a test stand for inertial sensing applications. Here we outline the current status of the experiment and report the observation of Ramsey interference fringes in the apparatus.}

\begin{document}
\maketitle
\flushbottom

\section{Introduction}
\label{sec:intro}

Atom interferometry is a high precision measurement technique [1]. Interference via atoms rather than light provides a theoretical 10$^{11}$ increase in sensitivity of gyroscopes [2], as well as achieving the world's most precise measurements of local gravity [3]. Atom interferometry can also, amongst other things, be used to measure Newton's constant [4], the fine-structure constant [5], to test Lorentz invariance [6], to test dark sector physics and as precision space-time sensors [7].

Here we present the initial performance from a drop-topology atom interferometer that has been developed for a search into the dark contents of the vacuum [8] and as a test stand for inertial sensing applications such as navigation and gravity scanning. Our current knowledge about the nature of the dark contents of the vacuum, such as dark energy, is entirely based upon cosmological observations. We intend to use atom interferometry as a possible probe of the dark contents of the vacuum on the laboratory scale [8].

An atom interferometer has been developed at low-cost by employing common-off-the-shelf (COTS) components with minor modifications using ultra-cold $^{85}$Rb as the atomic medium and a simplified two-laser optical system. Drawing on extensive experience and autonomy in complex radiation detection environments, bespoke and robust DAQ, control and detection systems have been developed for the apparatus. Using stimulated Raman transitions, interference fringes have been observed.

Development is underway with upgrades and the presented results influencing the decisions and directions for future improvements.

\section{Theory}
The ground states of alkali metals such as $^{85}$Rb are split into two hyperfine states which may be described using a semi-classical model of a two-level system [9]. Interaction of the atoms with coherent electromagnetic radiation allows the superposition of these states to be created and reliably controlled. Light-pulse atom interferometers measure the interference between the two states due to phase differences accrued by the light-atom interaction.

Light pulses with a frequency tuned to a two-photon transition between the states can coherently manipulate the state amplitudes. Varying the time the pulse impinges on the atoms causes the probability to be in each state to oscillate, a phenomenon known as Rabi oscillations. A pulse of characteristic duration, a `$\pi/2$' pulse, creates an equal superposition of the two states. A pulse of twice this length, a $\pi$ pulse, acts to invert the state population. By inserting a delay of time T between two $\pi/2$ pulses, the state populations change to produce interference fringes as the two-photon transition frequency is varied. These fringes are known as Ramsey fringes and are caused by a time dependent phase difference being accumulated between the pulses. This interference behaviour is only observable in systems which maintain coherence over the duration of the pulse sequence, requiring the atoms to be at extremely low temperatures.  For a more comprehensive description see [9].

\section{Method}

Initially $10^8$ $^{85}$Rb atoms are cooled within a magneto-optical trap (MOT) to achieve micro-Kelvin temperatures. An optical molasses is formed from which the atoms are released at a temperature of 15 $\mu$K. The temperature is limited by deliberately not cancelling the vertical component of the Earth's magnetic field. This field thereby acts as the axis of quantisation for the apparatus and separates the Zeeman substates. The atoms then fall freely under gravity to the interferometry region.

Two phase-locked laser beams with a stable frequency offset close to the hyperfine splitting n$_{HFS}$ = 3.0357 GHz [10] are required to coherently manipulate the state population. These drive stimulated Raman transitions between the F =2 and F = 3 hyperfine states of the $^{85}$Rb ground state.  These beams are generated using an acousto-optic modulator (AOM) which splits the carrier into two frequency components [11].

Control and DAQ for the Interferometer is a combination of a System on Chip (SoC) and FPGAs, allowing an interferometry sequence to be run repetitively over a long-time period. The system includes frequency generation for MHz range to control AOMs and can be chirped when necessary (providing compensation for the Doppler shift under the influence of gravity).

\begin{figure}[t]
\centering
\includegraphics[width=0.6\textwidth]{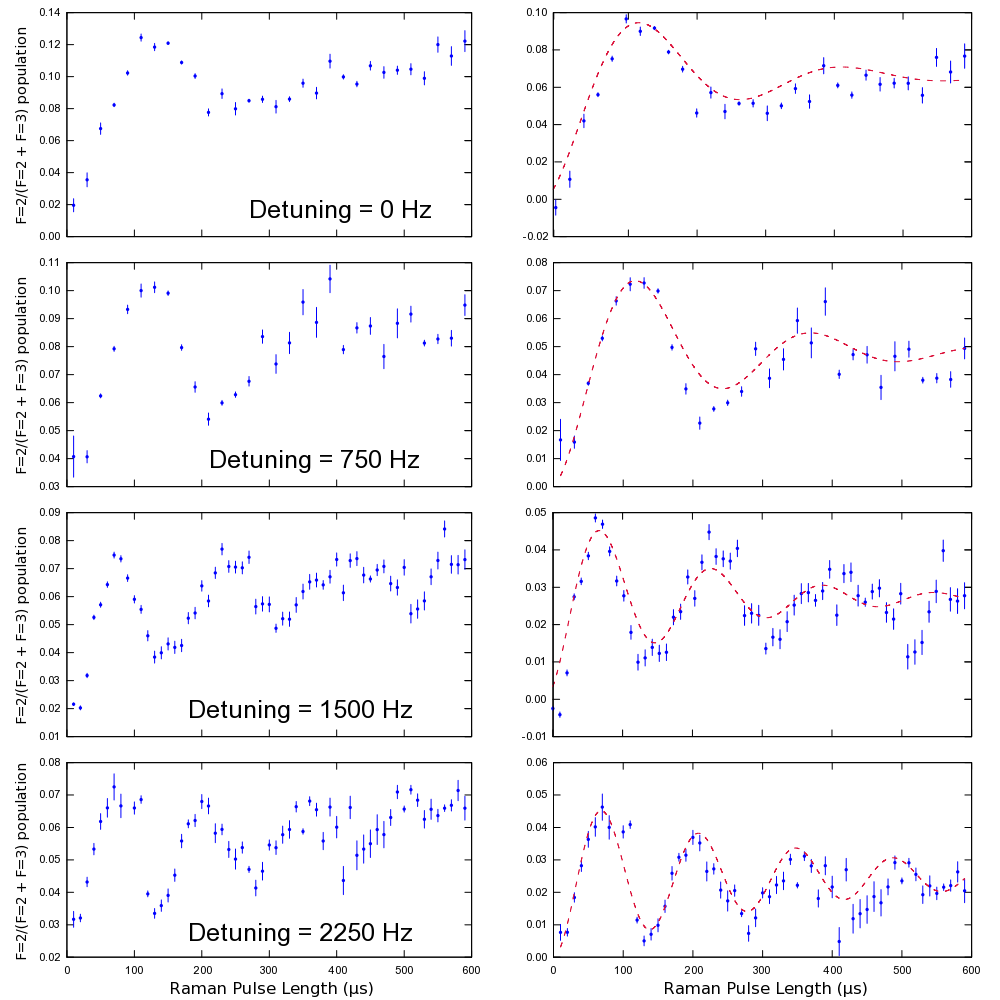}
\caption{Rabi oscillations for co-propagating Raman
beams. The plots on the left-hand side show the fractional population of the hyperfine ground states as function of
pulse length for different detunings. The plots on the right-hand side have a background subtraction applied. The
curve shows a fit to the expected oscillation function.
}
\end{figure}

\begin{figure}[b]
\centering
\includegraphics[width=0.7\textwidth]{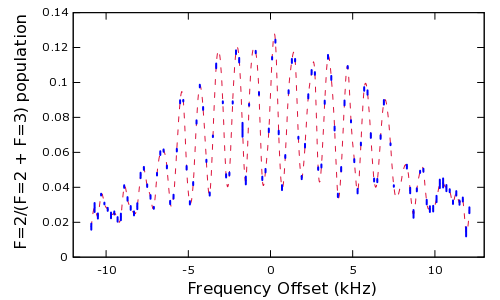}
\caption{Interference demonstrated through Ramsey
fringes, from a $\pi/2$ - $\pi/2$ pulse sequence (separation 800 $\mu$s). The dashed red line is a smooth function to guide the eye.}
\end{figure}

\begin{figure}[h]
\centering
\includegraphics[width=0.47\textwidth]{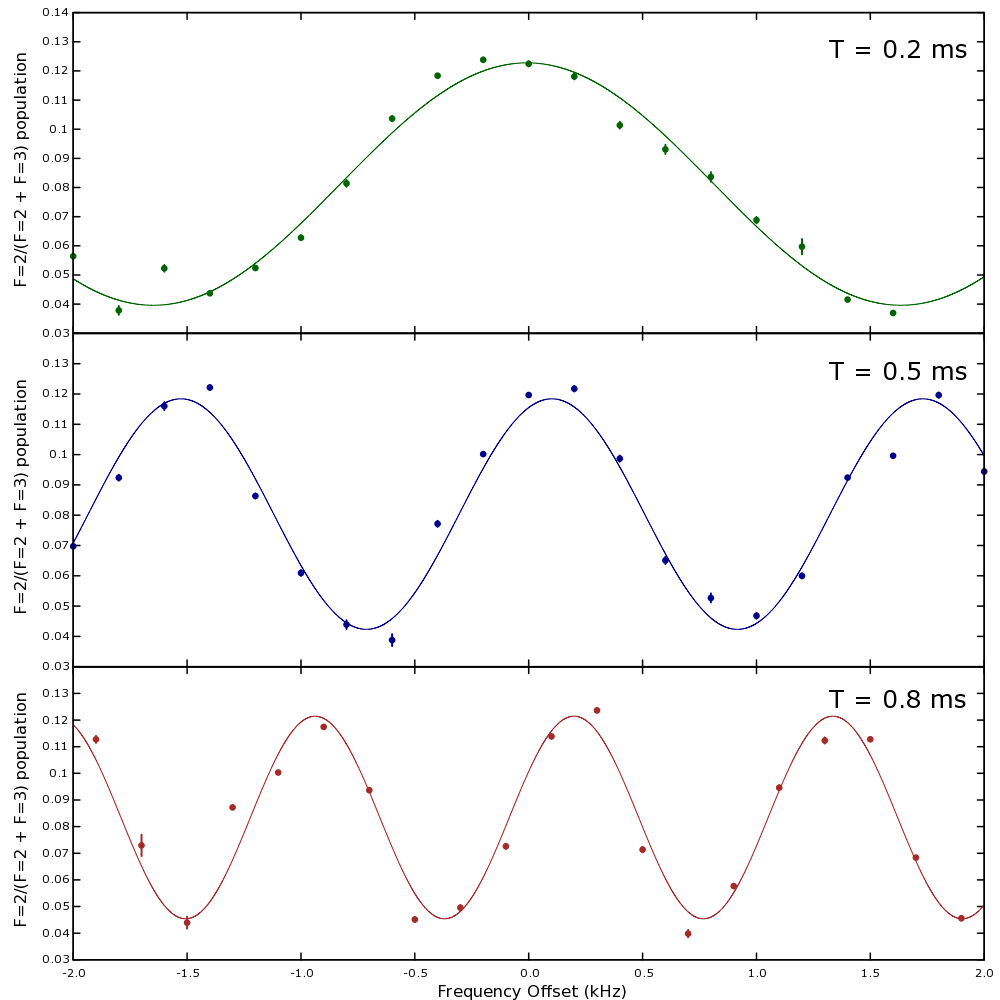}
\caption{Central interference fringes as a function of pulse
separation, fitted with a sinusoid.}
\end{figure}

\section{Results}

A search for Rabi oscillations was performed using co-propagating Raman beams. This was done by varying the total time of interrogation of the atoms whilst keeping the laser intensity constant.

The required frequency difference between the two Raman beams to provide maximum two-photon transitions was experimentally determined, with the results in figure 1 showing damped Rabi oscillations at various detunings from this resonance. The damping is due to far-from-resonance single-photon scattering causing decoherence and a linear increase in the state population. The effect of this is subtracted in the right-hand plots of figure 1. After 500 $\mu$s no oscillations are visible, indicating the limit of coherence of the present status of the Raman system. The data are fit with a sinusoid and an exponential damping term. For a detuning equal to zero, the $\pi/2$ pulse length is determined to be approximately 80 $\mu$s.

Adding the time delay T = 0.8 ms between two $\pi/2$ pulses demonstrates Ramsey interference fringes, as shown in figure 2. The central position is shifted by 200 Hz from the expected hyperfine splitting, consistent with an AC Stark shift.

Figure 3 shows the central fringe for three different pulse separations. The positions of the central peaks vary by approximately 200 Hz, less than 1 part in 10$^{7}$. The fringe height did not change as a function of pulse separation time, T, displaying no loss of coherence over these time periods.

\section{Summary}

We have developed a low-cost $^{85}$Rb atom interferometer, successfully realising the cold atom preparation, Raman population transfer, and high contrast interference fringes, all without the need for magnetic state-preparation. Upgrades are currently underway towards realization of the experiment's long term goals.

\section{Acknowledgements}

The authors would like to thank the support and encouragement of Themis Bowcock and the particle physics group at the University of Liverpool, STFC Daresbury Laboratory and the Cockcroft Institute, also the useful discussions with Yuri Ovchinnikov and the NPL. The work was supported by the Royal Society, AWE Ltd. We would like to thank the in particular Joseph Perl for their help and encouragement in this work.

\end{document}